\documentclass[12pt]{article}

\input{epsf.tex}

\begin{document}

\begin{center}
{\bf
APPROXIMATION OF SUMS OF EXPERIMENTAL RADIATIVE STRENGTH FUNCTIONS OF
DIPOLE GAMMA-TRANSITIONS IN THE REGION $E_\gamma \approx B_n$ FOR
THE ATOMIC MASSES $40 \leq A \leq 200$}\\
\end{center}\begin{center}
{A.M. Sukhovoj, W.I. Furman, V.A. Khitrov
 }\\
\end{center}\begin{center}
{\it Frank Laboratory of Neutron Physics, Joint Institute
for Nuclear Research, 141980, Dubna, Russia}\\
\end{center}

\begin{abstract}

The sums $k(E1)+k(M1)$ of radiative strength functions of dipole primary
gamma-transitions were approximated with high precision in the energy
region of $0.5 < E_1 < B_n-0.5$ MeV
for nuclei:
$^{40}$K, $^{60}$Co,
$^{71,74}$Ge, $^{80}$Br, $^{114}$Cd, $^{118}$Sn, $^{124,125}$Te, $^{128}$I,
$^{137,138,139}$Ba, $^{140}$La, $^{150}$Sm, $^{156,158}$Gd,
 $^{160}$Tb, $^{163,164,165}$Dy,  $^{166}$Ho, $^{168}$Er, $^{170}$Tm,
$^{174}$Yb, $^{176,177}$Lu, $^{181}$Hf, $^{182}$Ta, $^{183,184,185,187}$W,
$^{188,190,191,193}$Os, $^{192}$Ir,
$^{196}$Pt, $^{198}$Au, $^{200}$Hg
by sum of two
independent functions.
     It has been shown that this parameter of gamma-decay are determined by
the structure of the decaying
and excited levels, at least, up to the neutron binding energy.
\end{abstract}
\section{Introduction}
At present there are no doubts that the qualitatively differing
types of nuclear excitations co-exist, interact and have defining
influence on the structure and parameters of any nucleus.
Namely, they are quasiparticle and vibrational ones.
This is the main conclusion of such fundamental nuclear models as
QPNM \cite{QPNM} and different variants of IBM \cite{IBM}.

Unfortunately, the majority of experiments carried out by now gives
direct and quite reliable information on the structure of a nucleus
only for too small energies of its excitation.
Practically, for example, in \cite{W185} even-odd heavy nucleus this
region is up to now limited by the excitation energy interval of
$\approx 2$ MeV order.
Up to the present radiative strength functions of
$k=f/A^{2/3}=\Gamma_{\lambda f}/(E_{\gamma}^3\times
A^{2/3}\times D_{\lambda} )$ gamma-transitions in the whole range of
their energies $E_{\gamma}$, appearing at decay of any levels with
the excitation energies $E_i < B_n$, have been experimentally studied
to the least degree.

Main problems of the experiment, which occur at the determination
of $\rho$ and $k$ consist in the necessity to:

(a) really estimate the most serious systematical errors and to minimize
their values as much as possible;

(b) maximally exclude any model representations, which are unavoidable
at extracting the parameters of the process from the registered
spectra in an indirect experiment.

First of all, the latter refers to the basic hypothesis \cite{Bohr}
on the independence of the cross section of any reverse reaction from
the excitation energy of final nucleus used for the analysis of all
the experiments conducted up to the present.

According to particular calculations \cite{PEPAN-1996} of the
probabilities of gamma-transitions below $E_i \sim 3$ MeV and experimental
data \cite{PEPAN-2005} on the level cascade populating ability of
different types of nuclei below $0.5B_n$, for example,
it is expected that the hypothesis \cite{Bohr} is completely inapplicable
in even-even deformed nuclei. 

\section{The current state of experiment}

Between the decaying of compound states ${\lambda}$ and a group of the
low lying levels $f$ of the studied nucleus \cite{PEPAN-1991}  both
tasks of the experiment are solved to the most possible degree
by the registration of two-step cascade intensities $I_{\gamma\gamma}$
with the summed energy 5-10 MeV.

It is just the analysis of two-step cascade intensities of thermal neutrons
radiative capture in the fixed by \cite{Prim} $\Delta E$ energy intervals
of their intermediate levels $E_i=B_n-E_1$:
\begin{equation}
 I_{\gamma\gamma}(E_1)=\sum_{\lambda ,f}\sum_{i}\frac{\Gamma_{\lambda i}}
{\Gamma_{\lambda}}\frac{\Gamma_{if}}{\Gamma_i}
\end{equation}
that first revealed the
possibility of the model-free simultaneous determination of $\rho$ and
$k$ with their guarantee reliability.

In the initial variant \cite{Meth1} it was performed at assuming the
independence of partial radiative widths $\Gamma_{lm}$ of the gamma-transitions
of given energy and multipolarity from energies of any primary $l$ and
final $m$ levels (that is, using hypothesis \cite{Bohr}).
In contemporary version \cite{PEPAN-2005} it is accomplished without
its use in the region of the low lying levels ($E_i <0.5B_n$).

In order to practically calculate the cascade intensities and compare
them with the experiment it is necessary to take into account the real
specific character of the studied process.
In certain energy interval $\Delta E_j >> FWHM$ of the excited levels
there is only the value of $\Delta \Gamma_j$ sum of partial radiative
widths available for comparison with the calculation.
For any interval of number $j$ excitation energy this sum may be always
mathematically presented as a multiplication of a certain average of
partial width $<\Gamma_{j}>$  at $n_j=\rho \Delta E_j$ number of
excited levels. Equation (1) is transformed into 
\begin{equation}
I_{\gamma\gamma}(E_j)=
\sum_{\lambda ,f} 
\frac{<\Gamma_{j}>}{\sum_{k}<\Gamma_{k}> n_{k}} n_{j}
\frac{<\Gamma_{l}>}{\sum_m(<\Gamma_{m}> n_{m})}. 
\end{equation}
for any interval number $j$.

Since this sum includes partial widths of gamma-transitions between
levels of different structure (see, for example, \cite{PEPAN-1996}),
the average width of primary $<\Gamma_{l}>$ and secondary $<\Gamma_{m}>$
gamma-transitions is an averagely weighted value for any $j$ and $i$
intervals in the indicated representation.
It is determined by a particular ratio of quasiparticle and phonon
components of wave functions of the studied nucleus levels, belonging
to the $\Delta E$j interval.
It is necessary to take into account this circumstance when comparing
experimental $k$ values with theoretical representations.

System (2) includes $N$ of nonlinear equations and $2N$ of unknown
parameters. (In case, when the retio of radiative strength functions
of primary and secondary gamma-transitions is given on the basis of
an additional information \cite{PEPAN-2005}.) It describes a certain closed surface
in the space of $2N$ sought parameters.
Naturally, it is impossible to determine $\rho$ and $k$ unambiguous
values from (2) even when all available information on the considered
nucleus is used. However, in principle, it is always possible to determine
approximately or precisely the interval of variation of $\rho$ and $k$
values, setting given values of $I_{\gamma\gamma}$, $<\Gamma_{l}>$ and
other parameters of a particular nucleus.
Corresponding experiments have been carried out at the thermal neutron
capture for 51 and analyzed without using hypothesis \cite{Bohr} for\
22 nuclei.

\section{Main problems of determination and theoretical description
of parameters of cascade gamma-decay}

In \cite{PEPAN-2005,Meth1} techniques the maximum values of errors
$\delta \rho$ and  $\delta k$ of determined parameters are very strongly
limited by the type of spectra measured in the experiment.
Therefore, they always have quite acceptable value  \cite{TSC-err} for
practically obtainable systematical errors $\delta I_{\gamma\gamma}$
of measured distributions of cascade intensities.

It is just the obtained degree of accuracy of $\rho$ and $k$ values
determined by \cite{PEPAN-2005,Meth1} that defines both the reliability
of conclusion about factors determining the indicated nuclear parameters
and possibilities to extract new information on properties of particular
nuclei and, probably, on structure of initial cascade levels fixed by
conditions of the experiment.

For the most efficient extraction of data on the influence of nuclear
structure on $k(E1)+k(M1)$ and $\rho$ values from the experiment,
it is also necessary to use all the information accumulated by theory
on these parameters of nucleus.
Specifically, it is possible to separate two asymptotic variants of
behavior of the congruent radiative strength functions.
They appear when in an amplitude of gamma-transition dominate:

a) quasiparticle components of wave function of initial and final states or 

b) their one- or two-phonon components.

Such possibility follows from the successful description
\cite{PEPAN-2006,Prep196} of the experimental data for $\rho$ from
\cite{PEPAN-2005,Meth1} by the superposition of partial level densities,
determined by $n$-quasiparticle excitations with enhancement of
the quasiparticle level densities of $K_{vibr}>>1$ times at the phonon type
excitations
expense of the ofdomination.

Based on the main principles of fragmentation of nucleus compound-states
\cite{MalSol}, it is impossible to eliminate the probability of dominating
of either one or another component in the neutron resonance structure.
It is also impossible not to take into account the specific character
of excitations of various types.

Quasiparticle excitations (although with different number of
quasiparticles) are present at any excitation energy of nucleus.
Phonon ones, as it follows from the theoretical analysis of A.V. Ignatyuk
\cite{Kvib}, most probably become excited at energies close to those
of quadrupole (octopole) phonon and phonons of greater multipolarity.
In other words the excitations of phonon type may provide a significant
increase in radiative strength functions in the limited intervals of
nucleus excitation energy.
Therefore, the excitations of quasiparticle type may rather well determine
the basic part of radiative strength functions when gamma-transition
energy is changing (first of all -- primary), whereas monophonon
(multiphonon) one -- determine a position of local region and the degree
of increasing of strength functions in it.

Both variants \cite{PEPAN-2005,Meth1} of techniques use $k(E1)$ and $k(M1)$
unknown independent random functions to extract $\rho$ and $k$ from
$I_{\gamma\gamma}$ experimental values, as parameters of equation (2).
But near $E_\gamma=B_n$ (or smaller energy) they necessarily fix
$k(M1)/k(E1)$ ratio on the basis of the experimental data.
The same refers to the level densities of positive and negative parity.
Unfortunately, variations of the obtained random functions are rather
large. However, due to the strong anticorrelation of pairs of the listed
parameters, their sums fluctuate noticeably weak and, therefore,
are quite informative. Namely, these are well described by $\rho=\psi(E_{ex})$
and $k=\phi(E_1)$ functions, the parameters of which differ sufficiently
weak at changing of masses of the studied nuclei or when their changing
does not contradict the unconditionally set theoretical notions about
nucleus.

\section{Model for the semiphenomenological description of $k(E1)+k(M1)$}

All practically applicable \cite{RIPL} models of radiative strength
functions consider nucleus as a monocomponent object.
Owing to the historically prevalent representations, it is viewed as a
system of Fermi-particles.
Moreover, though model \cite{KMF} considers nucleus only as Fermi-liquid,
it reproduces a series of parameters of the cascade gamma-decay process
with better accuracy than is made possible by the simplest extrapolation
\cite{Axel} of cross section of reverse reaction into the region of
nucleus excitation below $B_n$.
By this reason, it is appropriate to use the representations \cite{KMF}
as basic in the performed analysis, because they are based on
the most realistic notion about nucleus produced by theorists by now.

A fraction of levels with purely dominating vibration components of
wave functions (at least lower than $\sim 0.5B_n$) is very significant
\cite{PEPAN-2006,Prep196}. However, the existing and practically used
\cite{RIPL} models of the radiative strength functions do not completely
take into account this fact.
Therefore, at this stage of the analysis it is necessary to postulate
the contribution of the vibrational nucleus excitations into $k(E1)+k(M1)$
values in a purely phenomenological way.

Below it is accepted that for a spherical nucleus fraction of partial
radiative width of dipole primary gamma-transitions with energy $E_{\gamma}$,
specified by the quasiparticle excitations, is described by model
\cite{KMF}:
\begin{equation}
k(E1,E_\gamma)+k(M1,E_\gamma)=w\frac{1}{3\pi^2\hbar^2c^2A^{2/3}} 
\frac{0.7\sigma_{\mathrm{G}}\Gamma_{\mathrm{G}}^2(E_\gamma^2+\kappa4\pi^2T^2)} 
{E_{\mathrm{G}}(E_\gamma^2-E_{\mathrm{G}}^2)^2},
\end{equation}
 with two free parameters: $\kappa$ and $w$.
For the deformed one -- by sum of two Lorenz curves, respectively.
Fraction $k(E1)+k(M1)$, supposedly connected with the vibrational
nucleus excitations, is approximated by peak, shape of which is preset by
two exponential functions: $Pexp(\alpha (E-E_p))$, $Pexp(\beta (E_p-E))$
for its low and high energy parts.
A resolved single peak (doublet or multiplet of peaks) is observed,
at least, in 30 out of 40 nuclei in the energy range of primary
transitions $E_p \sim 3-6$ MeV. In these $k(E1)+k(M1)$ and $\rho$ values
are simultaneously determined from cascade intensities with a moderate
systematical error.
Exponential dependence of form of its peaks successfully describes all
the data obtained by the present.

Parameter $\kappa$ in (3) takes into account a possible change of
nucleus thermodynamic temperature in form $T=\sqrt{\kappa U/a}$,
whereas parameter $w$ -- net effect of changing in retio of levels
fraction of quasiparticle and vibration types, both in the region of
neutron resonances, and intermediate ($E_i<B_n$) levels of nucleus.
Or it considers the fact of changing in ratio of quasiparticle
and vibtation components in the normalization of wave functions of
decaying and excited levels, respectively.
Such conclusion follows from the strong anticorrelation of amplitudes of
local peaks in the experimental data on $k(E1)+k(M1)$ and their
uninterrupted distribution (approximated by function (3)).

First of all the approximation of the experimental values of $k(E1)+k(M1)$
sums from \cite{PEPAN-2005} is used in the analysis.
When they are absent, those from \cite{Meth1} are used.
The results are presented in Fig.1-4 for the main part of the studied by
the present nuclei.
Particular values of the parameters by the approximating curve for
all nuclei \cite{PEPAN-2006,Prep196} are given in the Table.

Upon taking into account the type of compound state established by
theory of components of wave function and laws of nuclear states
fragmentation determining their value, the value of $w$ parameter is
to a greater or lesser extent determined by the structure of the wave
function of neutron resonance.
This implies that $w$ parameter may change from resonance to resonance
and, probably, correlate with values of their reduced neutron widths.

\section{Results of $k(E1)+k(M1)$ approximation}

Proposed for all studied nuclei the simplest form of functional dependence
of the sum of radiative strength functions provides rather accurate
reproduction of the experimental data \cite{PEPAN-2005} or \cite{Meth1}
in the energy interval of primary gamma-transitions $0.5 <E_1 <B_n-0.5$
MeV. For many of them it is sufficient to postulate the existence
of the single peak. However, when the experimental data for nuclei close
to magic ones is taken into account, most probably there are no less than
two local peaks in the radiative strength functions.
Positions of peaks well correspond to the region of step structure
in levels density (Fig.5).
This circumstance in an ensemble with a wide variety of values of the
adjusted parameter alpha in different nuclei is the basis for
interpretation of increasing of the strength function in regard to
model predicted value \cite{KMF} for those of $E_1 >1-3$ MeV.

The excitation energy, corresponding to the positions of local peaks in
the radiative strength functions of the considered nuclei (for $E_1>1$ MeV)
is compared with the threshold energy of appearing four 
(five)-quasiparticle levels in Fig.5.
Taking into account unavoidable errors and ambiguity of the analysis
\cite{PEPAN-2006}, their complete compliance with the position of one
of the peaks is observed. In the first approximation in the scale of
$B_n-E_1$ the position of another one coincides with the energy of
quadrupole phonons (for example, about 1 MeV for the deformed even-even
nuclei) or the region of mainly one- or two-particle levels in even-odd
and odd-odd nuclei, respectively.

Amplitudes of peaks, supposedly caused by phonon excitations, for the
most of analyzed nuclei are in a quite limited interval of values 
(Fig.6). Their noticeable increase is observed for $^{40}$K, $^{60}$Co,
$^{184}$W, $^{188,190}$Os, $^{192}$Ir  $^{200}$Hg.
That is for nuclei close to the double magic ones with nucleon number
$N(Z)$=20, 28 and for those in the region about $Z$=82 and $N$=126.

The best values of parameters $\kappa$ and $w$, as it may be seen from
Fig.6, obviously indicate smaller temperature of nucleus than the
thermodynamic ones even near $B_n$ and significant decrease of a
component $k(E1)+k(M1)$ by the approximating function (3).

Unfortunately, $\kappa$ and $w$ parameters very strongly correlate
both with one another and with $\alpha$ parameters in the region of the
smallest energies of primary gamma-transitions.
There are also no theoretical grounds to regard the form of local peaks
as the only possible in strength functions.
By this reason it is also impossible to exclude a possibility of
an additional systematical error of $\kappa$ and $w$ parameters.

In the proposed interpretation of the approximation data of radiative
strength functions appearing of the local peaks is postulated by the
presence of the amplified gamma-transitions on levels with the large
phonon components of wave functions.
If that is true, $\alpha$ parameter may be determined by change rate
of the nucleon correlation function in a heated nucleus.
In Fig.7 frequency distribution of this parameter is presented in nuclei
of different nucleon parity.
Additionally, frequency distribution of $\Delta_0^{-1}$ value is given
for even-even compound nuclei. (By positioning of maximum and width,
it is close to an analogous distribution $\Delta_0$ for nuclei with $A-2$).

The spread in $\alpha$ values is maximal for even-odd nuclei and
intermediate for odd-odd ones.
Assuming that $\alpha \propto \Delta_0^{-1}$, it is possible to determine
that two unpaired nucleons of the single type decrease $\Delta_0$ value
by 25-50\% at increasing of correlation function from one to three MeV.
A pair of neutron and proton quasiparticles decreases $\Delta_0$ somewhat
more intensely. Its most decrease is observed for even-odd nuclei.
In other words, in the region below the neutron binding energy the
radiative strength function depends on the type and number of nucleons
causing the appearance of excitations of quasiparticle type.

\section{Conclusion}

By the present the most reliable systematical experimental data on sums
of the radiative strength functions of dipole cascade gamma-transitions
of the thermal neutrons capture may be approximated in the frames of the
experiment unsertain in the $0.5 <E_1 <B_n-0.5$ MeV region by
a superposition of two components having rather different form of
the dependence on energy of gamma-transition.
At the present they may be interpreted only as a contribution of fermion
and boson nucleus excitations into the partial radiative width,
respectively.

Noticeable variation of the obtained parameters of $k(E1)+k(M1)=F(E_1)$ dependence in nuclei of different masses may be caused by strong influence of the structure of cascade initial level on the correlation of contributions of nuclear common and superfluid states into the probability of gamma-quanta emission. This effect appears as an addition to the observed in \cite{PEPAN-2005} strong influence of the intermediate level structure on this value.

Correspondingly, the radiative strength functions of primary
gamma-transitions of the decay of compound states may be determined by
the structure of levels connected by them at least up to the energy of
neutron binding.

{\sl Table~1.\\ } Parameters of approximation.
\begin{center}
\begin{tabular}{|r|l|l|l|l|l|l|}  \hline
Mass&  $\kappa$&  $w$  & $E_1$, MeV & $P, 10^{-9}$, MeV$^{-3}$  &
 $\alpha$, MeV$^{-1}$ & $\beta$, MeV$^{-1}$  \\\hline
 71&-0.01(37)& 0.004(1)& 5.49(7)& 1.07(9)&  0.86(10)& 2.07(46)\\
   &        &         &  6.52(5)& 1.24(7)&  2.06(43)& 0.22(18)\\
   &        &         &  0.77(11)&0.46(11)& 0(17)&    3.2(15)\\
125& 0.42(8)&  0.35(4)&  4.79(9)& 0.83(10)& 0.52(10)& 2.4(8)\\
   &        &         &  6.02(9)& 0.89(20)& 2.56(84)&-0.2(51)\\
137& 0.26(2)&  3.6(1)&   1.84(6)& 3.98(27)& 0.74(10)& 1.85(27)\\
   &        &         &  4.72(1)& 42.7(9)&  29(2)&    9.28(26)\\
139& 1.0(3 )&  0.5(1)&   3.94(9)& 1.99(29)& 2.81(85)& 0.10(42)\\
   &        &         &  0.50(39)&0.8(6)&   1.8(18)&  2.0(21)\\
163& 0.002(9)& 0.04(56)& 3.9(4)&  1.1(16)&  1(25)&    0.7(11)\\
   &        &         &  2.9(7)&  0.8(11)&  1.5(13)&  0.7(30)\\
   &        &         &  5.4(20)& 1.0(77)&  0.9(78)& -1.4(44)\\
165&-0.00(10)& 0.05(78)& 3.56(6)& 7.6(8)&   2.04(31)& 0.68(27)\\
181& 0.06(5)&  0.62(9)&  3.8(1)&  0.99(20)& 1.07(31)& 3.2(12)\\
   &        &         &  2.95(27)&0.39(15)& 1.38(95)& 1.04(69)\\
183& 0.17(13)& 0.04(1)&  2.9(1)&  0.72(7)&  0.77(13)& 1.00(22)\\
   &        &         &  5.37(1)& 8.83(18)& 3.49(12)& 0.66(37)\\
185& 0.29(5)&  0.45(3)&  3.23(3)& 2.27(7)&  1.55(10)& 0.80(4)\\
   &        &         &  4.74(1)& 2.09(7)&  3.58(35)& 1.43(32)\\
187& 0.23(4)&  1.03(5)&  3.15(7)& 1.08(9)&  1.19(20)& 1.16(16)\\
   &        &         &  5.14(9)& 1.05(46)& 5.1(15)&  1(86)\\
191& 0(77)&   -0.0(30)&  4.3(2)&  2.2(14)&  2(15)&    1(11)\\
   &        &         &  3.8(2)&  2.3(11)&  2.2(83)&  1(12)\\
193& -0.1(10)& 0.00(1)&  3.3(2)&  2.0(46)&  2.2(45)&  0.4(48)\\
   &        &         &  3.9(3)&  1.9(78)&  1.7(84)&  0.7(71)\\\hline
177& 0.30(7)&  0.43(8)&  6.0(1)&  3.71(28)& 0.56(7)&  0.44(30)\\
   &        &         &  4.0(2)&  2.59(19)& 0.63(10)& 0.00(5)\\\hline
40& -0.01(11)& 0.07(2)&  5.33(1)& 21.2(3)&  1.48(4)&  0.84(12)\\
   &        &         &  4.22(3)& 7.24(33)& 1.65(9)&  1.23(11)\\
 60& -0.02(3)& 0.8(5)&   6.01(6)& 12.8(6)&  0.41(2)&  1.63(37)\\
   &        &         &  6.80(2)& 17.9(20)& 6.0(13)&  1.6(10)\\
 80& 0.43(3)&  0.75(6)&  5.17(6)& 3.84(34)& 1.11(17)& 2.(4)\\
   &        &         &  7.42(2)&10.38(63)& 3.76(30)& 0.10(24)\\
   &        &         &  0.52(8)& 4.3(15)&  1(90)&    4.2(14)\\\hline
\end{tabular}\end{center}
\begin{center}
\begin{tabular}{|r|l|l|l|l|l|l|}
\hline
Mass&  $\kappa$&  $w$  & $E_1$, MeV & $P, 10^{-9}$, MeV$^{-3}$  &
 $\alpha$, MeV$^{-1}$ & $\beta$, MeV$^{-1}$  \\\hline
128& 0.05(2)&  0.66(9)&  4.7(3)&  0.91(20)& 0.61(20)& 0.79(46)\\
   &        &         &  5.6(2)&  1.41(20)& 0.50(10)& 1.01(45)\\
140& 0.07(15)& -0.02(4)& 3.77(4)& 1.28(9)&  1.32(15)& 2.42(40)\\
   &        &         &  4.84(2)& 3.27(18)& 7.05(83)& 1.46(30)\\
160& 0.29(2)&  1.20(7)&  4.16(6)& 5.05(33)& 1.09(13)& 0.99(13)\\
   &        &         &  5.93(5)& 8.91(30)& 0.69(4)&  0.49(41)\\
166& 0.15(6)&  0.60(16)& 4.12(4)&13.3(7)&   1.21(8)&  0.98(9)\\
   &        &         &  5.89(9)& 7.1(6)&   1.30(24)& 0.08(56)\\
170& 0.19(6)&  0.78(11)& 3.5(2)&  1.53(31)& 0.71(30)& 1.6(9)\\
   &        &         &  6.45(6)& 8.89(38)& 0.89(8)&  0.01(12)\\
176& 0.0(55)&  0.5(19)&  3.9(1)&  9.6(60)&  1.86(83)& 1.0(14)\\
   &        &         &  4.85(14)&6.1(45)&  1.8(41)&  3(5)\\
182& 0.14(9)&  0.14(7)&  4.5(1)&  1.50(23)& 0.85(22)& 1.8(7)\\
   &        &         &  6.2(11)& 0.37(25)& 0.58(60)& 1(51)\\
192& 0.11(1)&  2.92(13)& 5.57(5)& 14.5(5)&  0.61(2)&  1(5)\\
198& 0.09(2)&  2.62(11)& 4.99(9)& 2.23(33)& 0.87(17)& 2.16(75)\\
   &        &         &  5.8(1)&  1.68(33)& 1.04(36)& 2.6(19)\\\hline
74& 0.14(6)&  0.25(4)&  5.37(6)& 3.03(12)& 0.59(35)& 0.58(4)\\
   &        &         &  6.5(1)&  1.81(10)& 1.20(24)& 0.33(7)\\
114& 0.18(9)&  0.10(4)&  5.68(6)& 4.06(25)& 0.89(6)&  1.11(20)\\
   &        &         &  7.68(15)&4.84(86)& 1.16(16)& 1(85)\\
118& 0.04(25)& 0.01(4)&  5.04(7)& 3.36(24)& 0.90(8)&  0.95(16)\\
   &        &         &  7.94(7)& 10.3(7)&  0.87(4)&  1(50) \\
124& 0.18(3)&  0.56(6)&  7.3(1)&  2.87(32)& 0.80(9)&  0(98)\\
138& 0.12(6)&  0.42(13)& 6.1(5)&  0.85(59)& 1.28(85)& 2(20) \\
150& 0.09(5)&  0.22(4)&  4.81(8)& 2.45(17)& 0.80(8)&  0.89(13)\\
   &        &         &  7.3(2)&  1.27(33)& 1.38(38)& 0(91)\\
156& 0.28(14)& 0.19(3)&  5.56(5)& 3.33(9)&  0.73(3)&  0.26(4)\\
   &        &         &  7.20(1)&    14.4(3)&  52(26)&   0.20(7)\\
158& 0.20(7)&  0.41(8)&  5.1(2)&  2.71(18)& 0.42(6)&  0.01(7)\\
   &        &         &  6.85(2)& 10.6(5)&  5.38(62)& 0.02(57)\\
   &        &         &  0.2(17)& 0.40(30)& 1(50)&    0.43(42)\\
168& 0.002(3)& 0.27(36)& 5.1(4)&  1.2(8)&   0.66(79)& 0.7(12)\\
   &        &         &  4.1(7)&  0.7(7)&   0.59(82)& 0.7(18)\\
164& 0.02(8)&  0.29(18)& 5.3(1)&  5.8(7)&   0.98(15)& 0.98(23)\\
   &        &         &  7.2(2)&  3.3(8)&   1.42(71)& 0(75)\\
174& 0.08(9)&  0.16(9)&  5.0(3)&  1.4(2)&   0.58(15)& 0.32(22)\\
   &        &         &  6.1(1)&  1.6(3)&   1.7(7)&  -0.12(26)\\
184& 0.1(48)& -0.01(23)& 5.11(3)& 29.7(7)&  0.93(2)&  0.97(7)\\
   &        &         &  6.20(1)&  48(2)&    53(76)&   0.06(24)\\
188&-0.001(3)&-0.06(9)&  5.55(2)& 24.8(7)&  1.45(6)&  1.24(8)\\
190& 0.09(3)&  0.53(10)& 5.18(6)& 14.8(8)&  0.75(6)&  0.57(6)\\
   &        &         &  6.47(4)& 20.3(7)&  1.22(13)& 0.47(5)\\
196& 0.02(7)&  0.04(5)&  5.34(6)& 4.2(3)&   1.09(12)& 0.89(13)\\
   &        &         &  6.57(2)& 6.0(38)&  16.4(79)& 44(42)\\
200&-0.02(8)& 0.04(2)&   4.95(1)& 11.7(1)&  1.39(2)&  1.08(2)\\
   &        &         &  5.97(1)&  9.0(2)&  1.56(5)&  7.0(3)\\\hline
\end{tabular}\end{center}

\newpage
\begin{figure}
[tbp]
\vspace{-5cm}
\begin{center}
\leavevmode
\epsfxsize=15.cm
\epsfbox{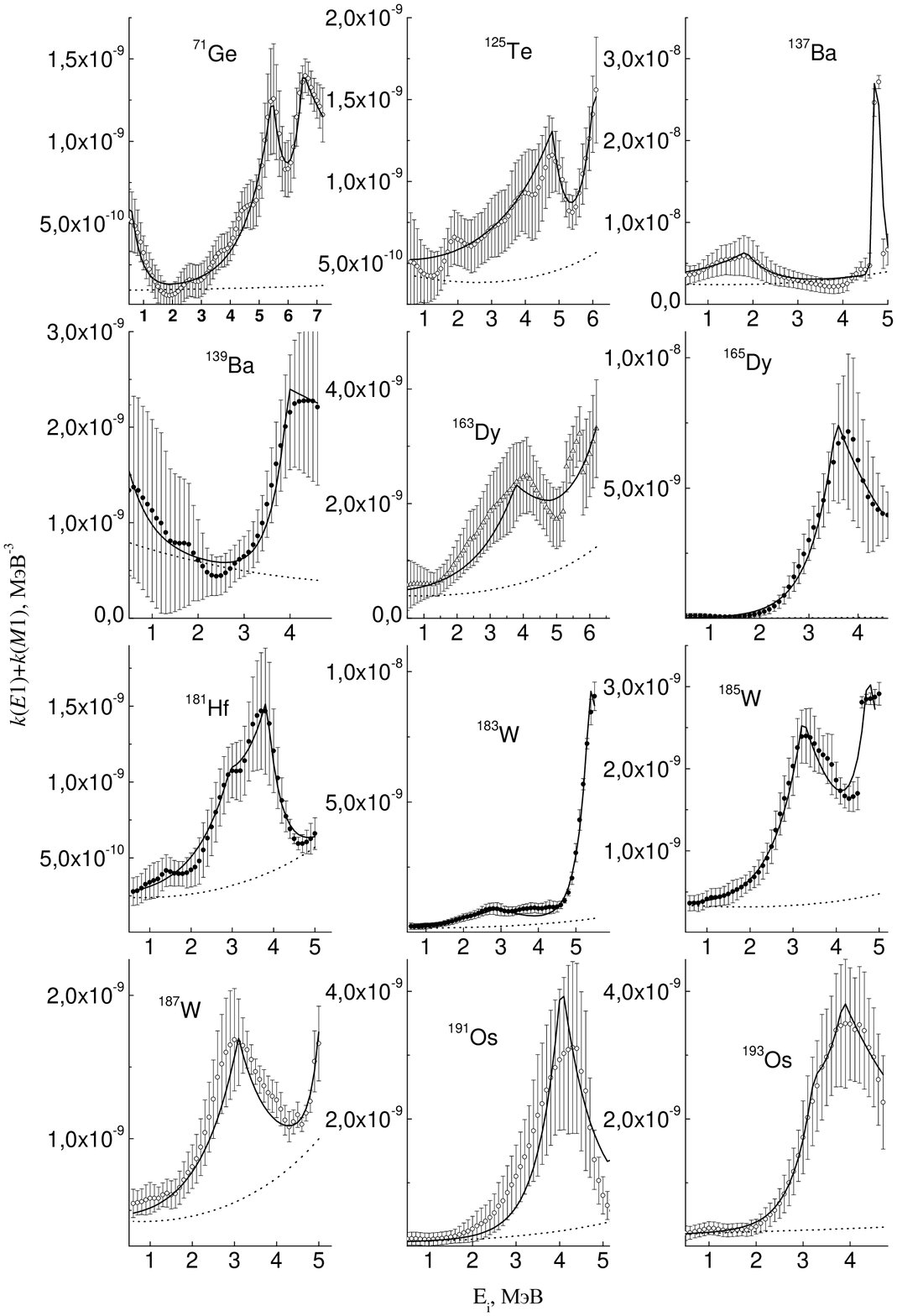} 

\end{center}
\vspace{0cm}
{\bf Fig.~1.} Points with errors --the most probable values of sums of radiative
strength functions and interval of their values, corresponding to
minimum values $\chi^2$ for even-odd compound nuclei.
Open points -- data from \cite{Meth1}, full ones -- from
\cite{PEPAN-2005}.
Line -- the best fitting.
Dots - component corresponding to equation (3).

\end{figure}

\newpage
\begin{figure}
\vspace{-5cm}
\begin{center}
\leavevmode
\epsfxsize=15.cm
\epsfbox{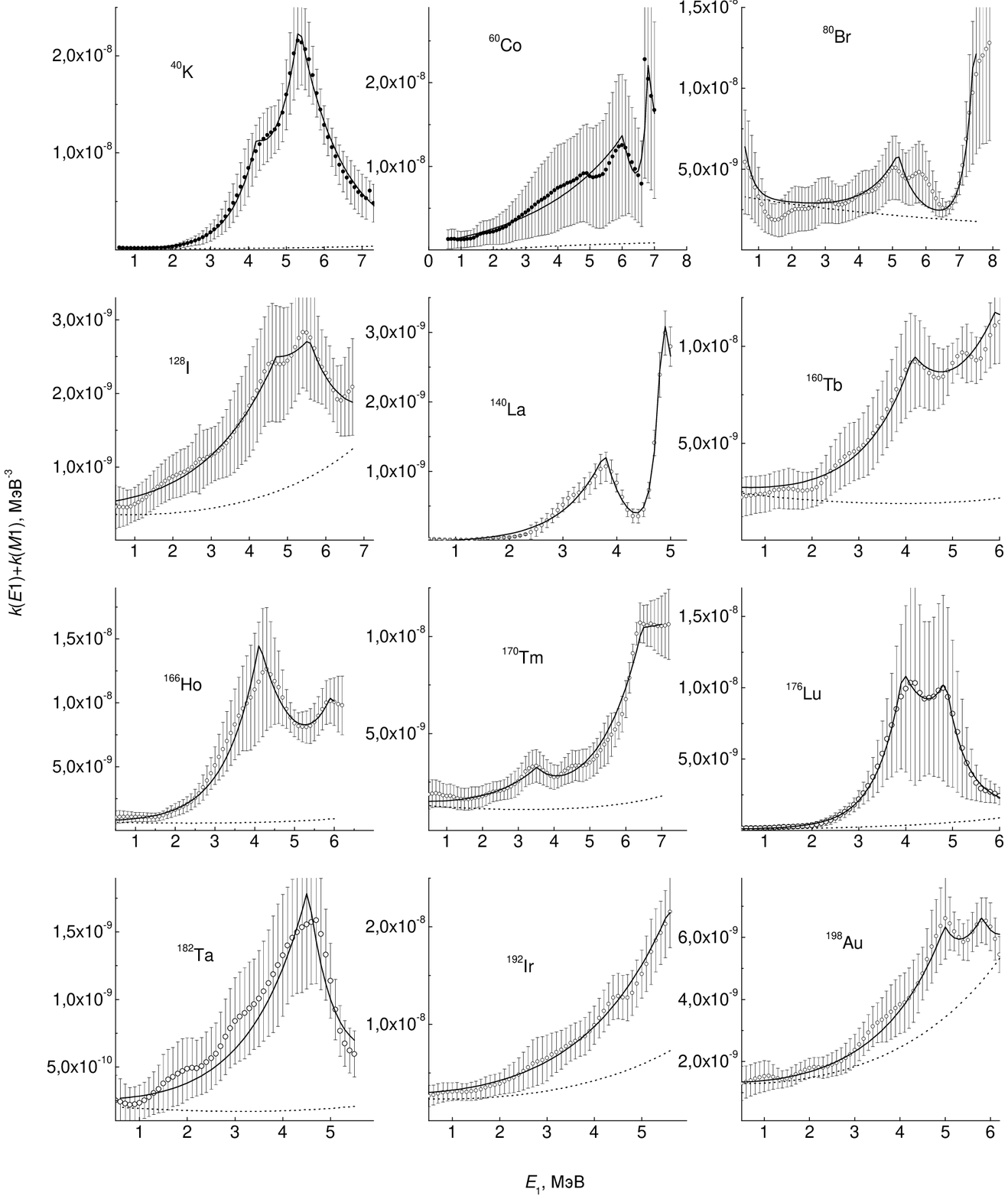} 

\end{center}
\vspace{-4cm}
{\bf Fig.~2.} The same as in Fig.1 for odd-odd nuclei.

\end{figure}

\newpage
\begin{figure}
\vspace{-3cm}
\begin{center}
\leavevmode
\epsfxsize=15.cm
\epsfbox{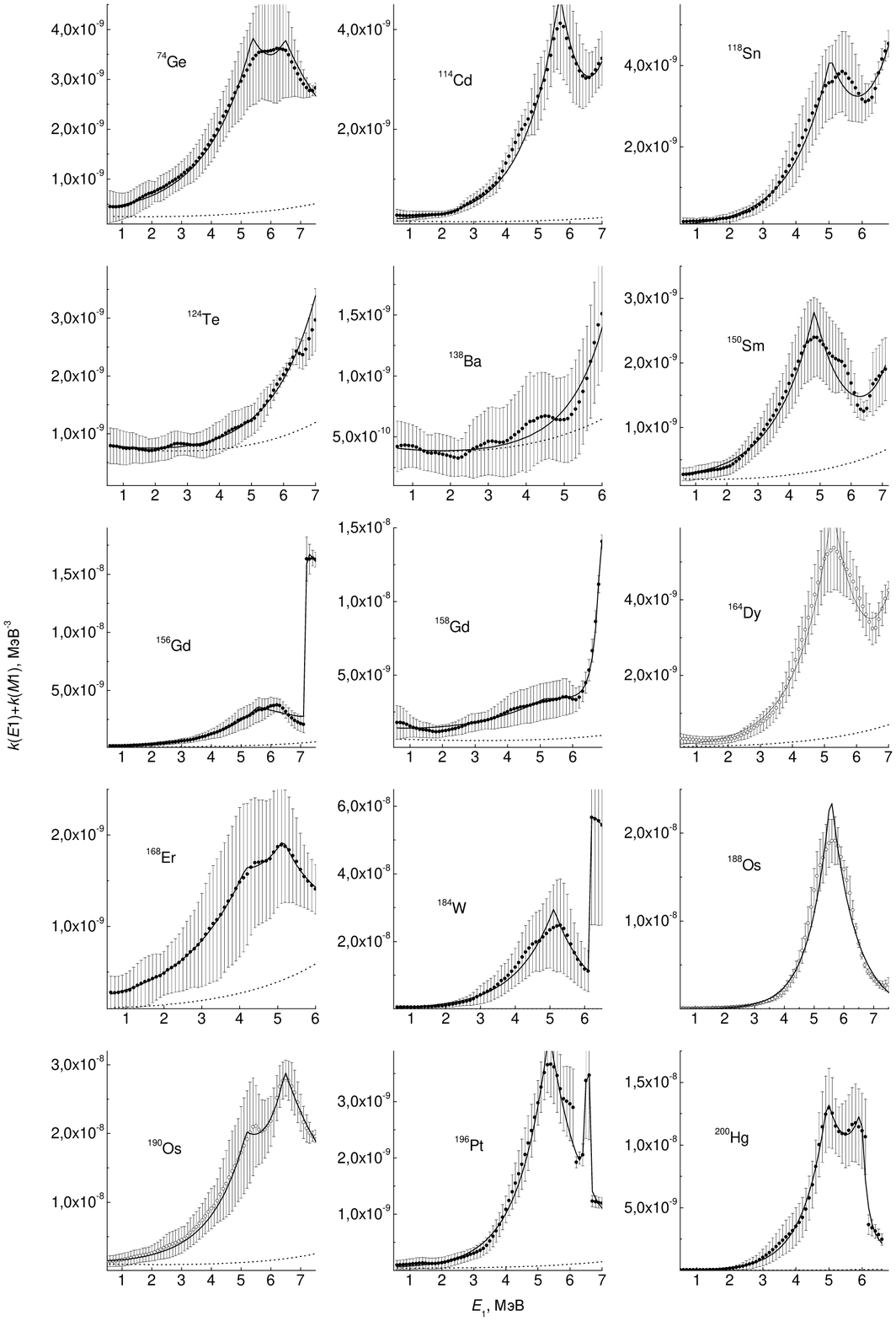} 

\end{center}

{\bf Fig.~3.} The same as in Fig.1 for even-even nuclei.
  
\end{figure}
\newpage
\begin{figure}[t]
\begin{center}
\leavevmode
\hspace{-2.8cm}
\epsfxsize=10.5cm
\epsfbox{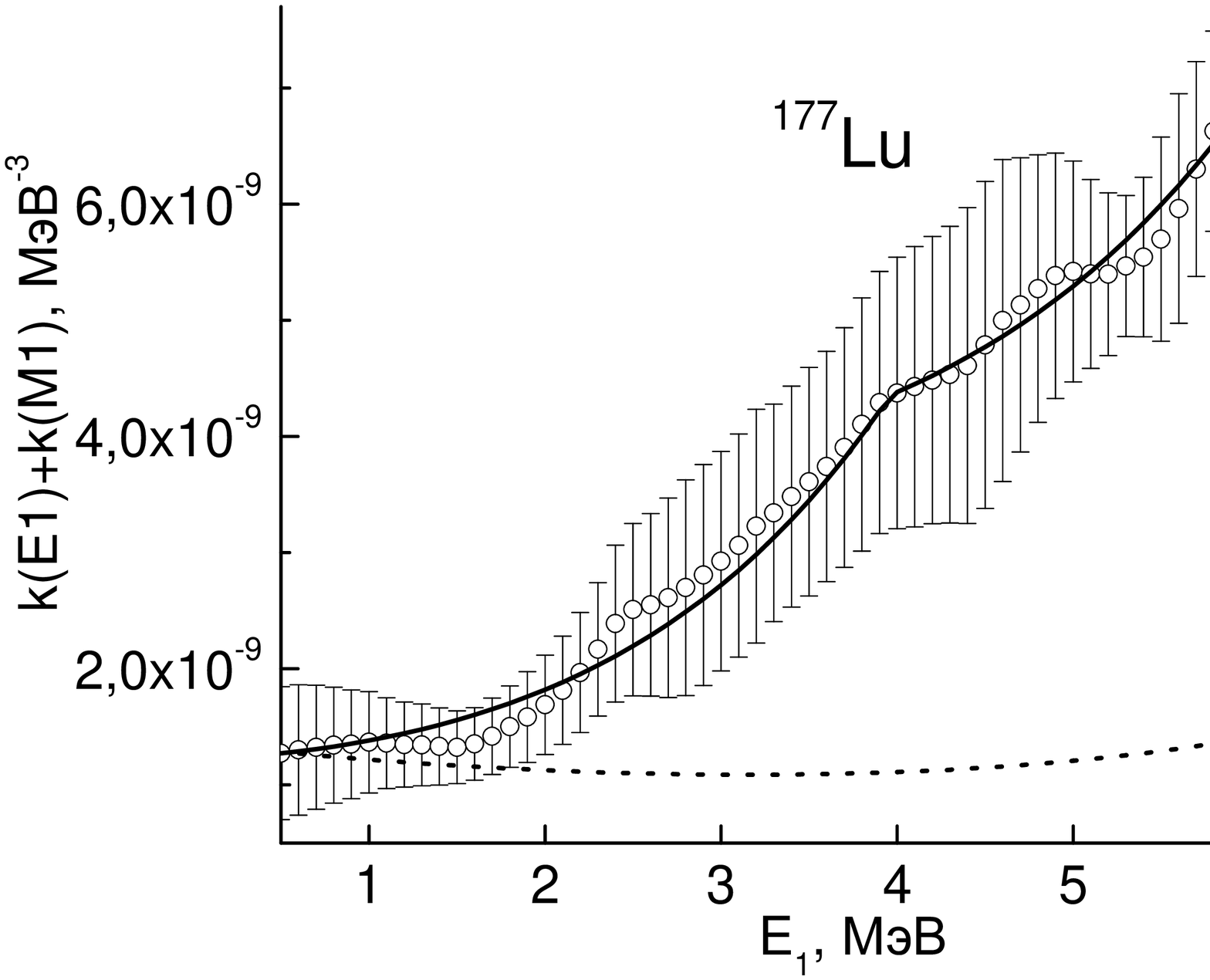} 

\end{center}
\vspace{-2cm}
{\bf Fig.~4.}  The same as in Fig.1 for $^{177}$Lu odd-even nucleus.

\end{figure}
\newpage
\begin{figure}
[t]
\vspace{-2cm}
\begin{center}
\leavevmode
\hspace{-2.8cm}
\epsfxsize=16.5cm
\epsfbox{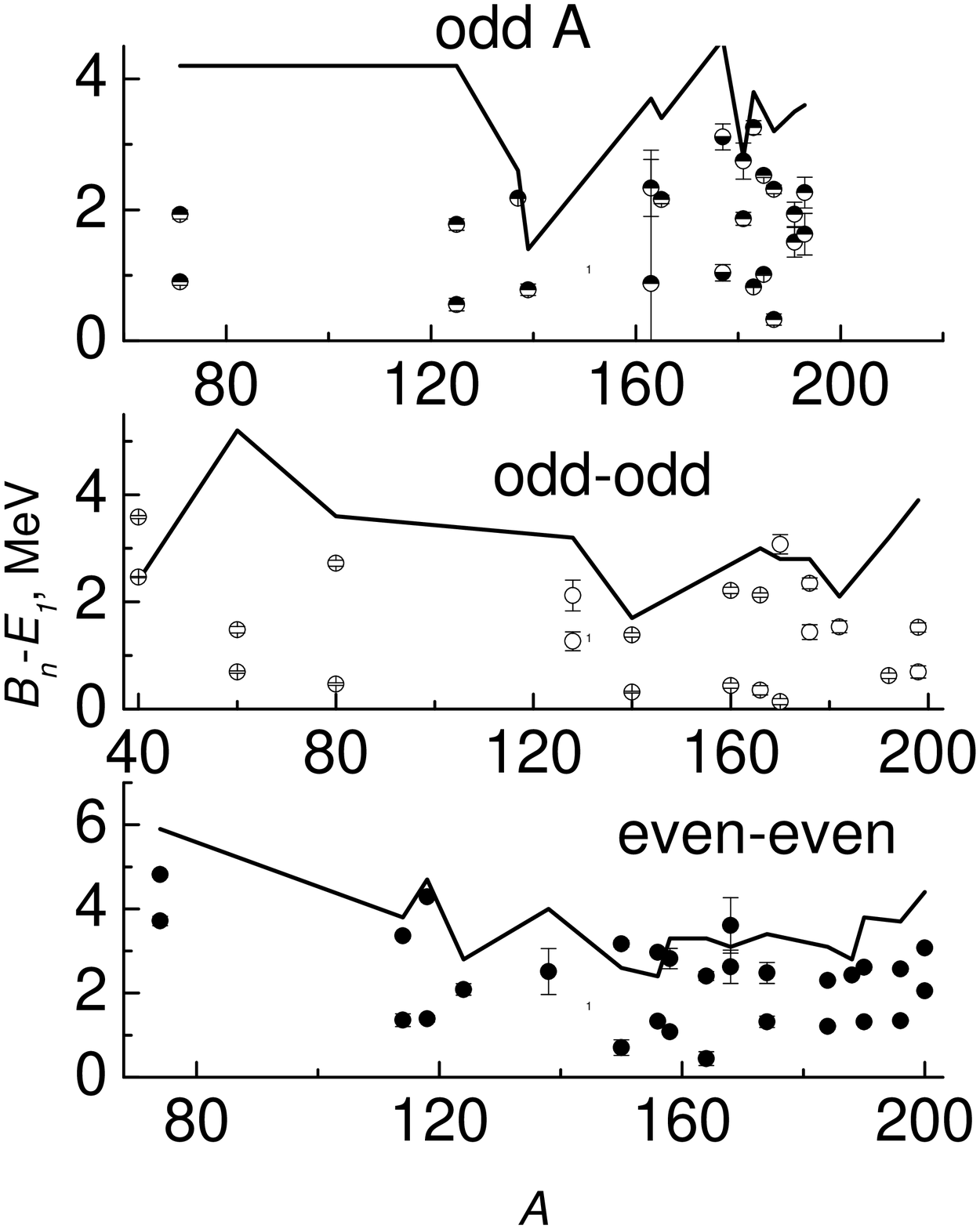} 

\end{center}
\vspace{-1cm}
{\bf Fig.~5.} Excitation energy of studied nuclei corresponding to positions
of local fluctuations (maxima) $k(E1)+k(M1)$ -- points.
Broken line -- threshold of the excitation of four- and five
quasiparticles in A-even and A-odd nuclei, respectively, for variant
of approximation from \cite{PEPAN-2006}.
Data for $^{177}$Lu are registered with those for even-odd nuclei.

\end{figure}
\newpage
\begin{figure}
\vspace{-4cm}
\begin{center}
\leavevmode
\hspace{-2.8cm}
\epsfxsize=14.5cm
\epsfbox{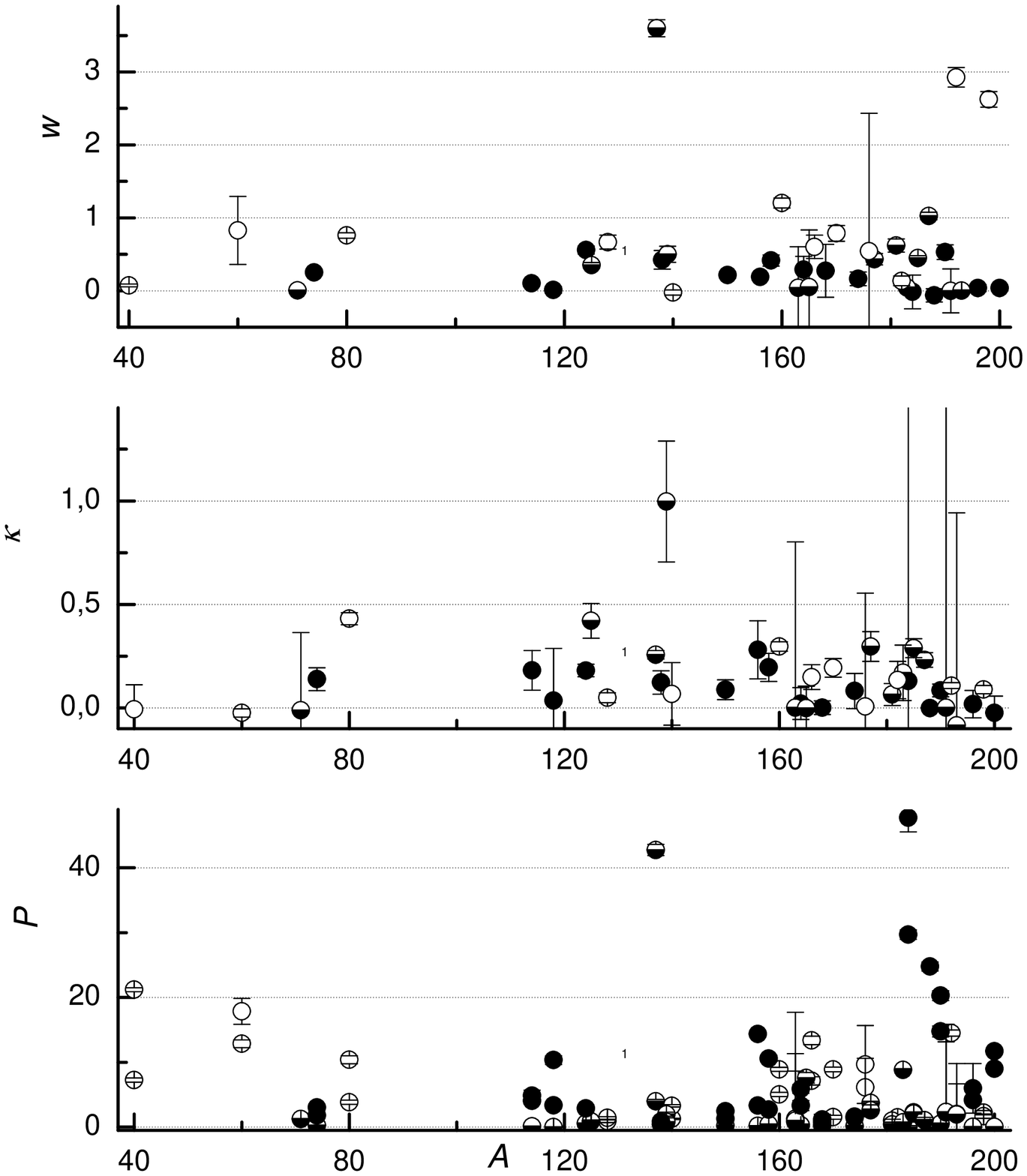} 

\end{center}
{\bf Fig.~6.} Amplitudes $P$ of local peaks (multiplied by $10^9$). Values
of $\kappa$ and $w$ parameters for function (3).
Labels of nucleon number parity are analogous to those in Fig.5.

\end{figure}
\begin{figure}
\vspace{-5cm}
\begin{center}
\leavevmode
\hspace{-2.8cm}
\epsfxsize=14.5cm
\epsfbox{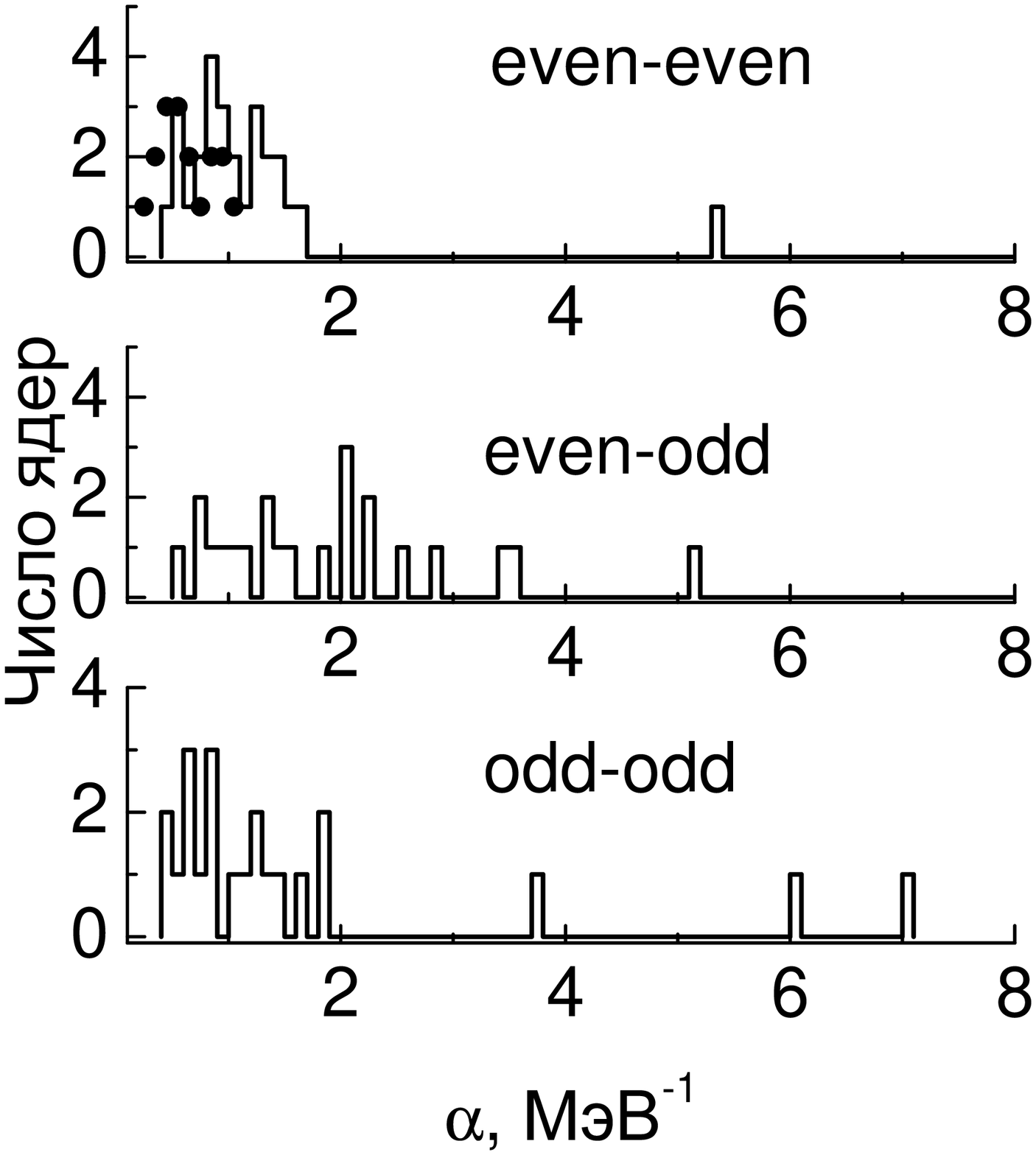} 

\end{center}
\vspace{-2cm}
{\bf Fig.~7.} Histogram -- frequency distribution of $\alpha$ parameter in
nuclei with different nucleon number parity.
Points -- the same for $\Delta_0^{-1}$.
\end{figure}

\end{document}